%% file: main.tex
\documentclass[preprint,12pt]{elsarticle}

\usepackage{amssymb}
\usepackage{amsmath}
\usepackage{gensymb}
\usepackage{bm}
\usepackage{graphicx}
\usepackage{caption}
\usepackage{booktabs}
\usepackage{url} 
\usepackage{mathtools} 

\usepackage{IEEEtrantools}
\usepackage{newtxmath} 
\usepackage{physics} 
\AtBeginDocument{\RenewCommandCopy\qty\SI} 
\usepackage{float}
\DeclareMathAlphabet{\mathcal}{OMS}{cmsy}{m}{n} 

\usepackage{subcaption}

\usepackage{xcolor}
\usepackage[linesnumbered,ruled,vlined]{algorithm2e}

\SetCommentSty{mycommfont}
\SetKwInput{KwInput}{Input}                
\SetKwInput{KwOutput}{Output}              
\journal{Digital Signal Processing}

\allowdisplaybreaks

\begin{document}

\begin{frontmatter}



\title{An All-Digital Coherent AFSK Demodulator for CubeSat Applications} 


\author[UFC]{Rubem Vasconcelos Pacelli} 
\author[ITA]{Rodrigo de Lima Florindo} 
\author[ITA]{Felix Antreich} 
\author[LUMINA]{Antônio Macilio Pereira de Lucena} 


\affiliation[UFC]{organization={Department of Teleinformatics Engineering (DETI), Federal University of Ceará (UFC)},
addressline={Campus do Pici - Bloco 725}, 
city={Fortaleza},
postcode={60455-970}, 
state={Ceará},
country={Brazil}}



\affiliation[ITA]{organization={Aeronautics Institute of Technology (ITA)},
addressline={Praça Marechal Eduardo Gomes, 50}, 
city={São José dos Campos},
postcode={12228-900}, 
state={São Paulo},
country={Brazil}}

\affiliation[LUMINA]{organization={Lumina Institute of Research, Innovation and Applied Technology},
  addressline={Avenida Santos Dumont, 1699.}, 
  postcode={60150-161}, 
  city={Fortaleza},
  state={Ceará},
  country={Brazil}}

\begin{abstract}
\input{secs/abstract.tex}
\end{abstract}




\end{frontmatter}



\input{secs/intro.tex}
\input{secs/channel.tex}
\input{secs/demodulator.tex}
\input{secs/results.tex}
\input{secs/conclusion.tex}

\section*{Acknowledgements}

This work was supported by the Conselho Nacional de Desenvolvimento Científico e Tecnológico (CNPq) under grant 312394/2021-7 PQ-2, 406517/2022-3, and 165074/2021-3, as well as Coordenação de Aperfeiçoamento de Pessoal de Nível Superior (CAPES) under the grant 88887.975685/2024-00.





\bibliographystyle{elsarticle-num-names} 
\bibliography{refs_modified.bib}



\end{document}

%% file: secs/abstract.tex
Audio frequency-shift keying (AFSK) is a widely adopted modulation scheme for CubeSat systems due to its favorable bandwidth efficiency and implementation simplicity. However, coherent detection usually is avoided because synchronization impairments, caused by intense line-of-sight (LOS) dynamics inherent in low Earth orbit (LEO), may significantly degrade the bit error rate (BER). This paper presents a new all-digital coherent AFSK demodulator based on a Kalman filter (KF) for carrier phase and timing delay synchronization and the Viterbi algorithm for bit detection. The Viterbi algorithm is employed for maximum likelihood sequence detection, and the detected bit statistics are fed back to the KF to estimate phase shift, Doppler frequency shift, and Doppler drift induced by the LOS dynamics. Original mathematical analyses are derived to provide a theoretical foundation for the proposed demodulator's operation, specifically addressing its synchronization accuracy in dynamic LEO environments. The proposed demodulator is evaluated considering an additive white Gaussian noise channel with real CubeSat orbits. The performance results obtained through computer simulations demonstrate that the proposed model can withstand such scenarios with a gain of \(5\) dB in terms of BER compared to the conventional noncoherent AFSK demodulator. The KF performance is assessed using a moving root-mean-square error (MRMSE) statistic and the trace of its state error covariance matrix estimate.

%% file: secs/intro.tex
\section{Introduction}\label{sec1}

A CubeSat is a class of small-scale satellites with a standard unit defined as a 10 cm (3.9 in) cube weighing up to 2 kg. Its fast development time, small dimensions, and lightweight were the main reasons for making space exploration affordable for universities, small- and medium-sized companies, and non-governmental organizations. On the one hand, CubeSats significantly contributed to the exponential growth of the number of satellites launched in the last two decades, which fosters opportunities for innovation and advancements in the satellite market technology \citep{emanuelliaNanosatellitesTheirDemand2016}. On the other hand, the overcrowding of radio bands allocated for satellite communication (mainly in low orbit) has concerned engineers and scientists, who have proposed new spectrally efficient modulations as alternatives to the current schemes \citep{gaysinSurveyModulationCoding2017}.

Despite the variety of modulations adopted in recent years, it is possible to notice a consistency in those used in service telecommunications subsystems \cite{CubeSatCommunicationsSystem}. These subsystems of the satellite are indispensable to any space mission as they establish telemetry, tracking, and telecommand communication with the control station. Its main functions are \cite{gerasenkoBeaconSignalsWhat2001, noeDesignImplementationCommunications2004}:
\begin{enumerate}
  \item Receiving and retransmitting tracking signals to enable measurements of the ground-satellite distance and the radial velocity, allowing to determine its geographical location and its orbit parameters
  \item Receiving telecommand signals from the control station to initiate manoeuvres and to change the satellite operation
  \item Sending telemetry signals generated by the on-board computer during the housekeeping (beacon signals). This data contains information about the satellite's health, such as temperature operation and voltage
\end{enumerate}
The main on-board equipment of service telecommunication subsystems is the telemetry, tracking, and command (TT\&C)  module, which comprises a set of antennas, an analogue front-end, and a digital modem. Its functions are modulating and demodulating the transmitted and received signal, respectively. Audio-frequency shift keying (AFSK) is widely used for the uplink of telecommand signals due to its compatibility with the data layer used in most CubeSat applications, the AX.25 protocol \cite{beechAX25LinkAccess1998}. This allows amateur radio operators worldwide to collect information from the satellites and, in some cases, improve global availability. Simplicity is also a decisive factor in the widespread use of AFSK. In general, inexpensive programmable devices can be used to implement the AFSK modulator \cite{velascoSoftwareBasedAFSK2018}.

On the modulator side, the AFSK signal is usually implemented with a continuous phase. Due to its constant envelope property, the AFSK signal allows using nonlinear amplifiers without causing spectral regrowth \cite{zhouAnalysisSpectralRegrowth2000}. Such devices are 2 dB to 3 dB more efficient than power amplifiers of class A or AB \cite{andersonAdvancesConstantEnvelope1991}. Concerning the demodulator, the literature reports noncoherent detection as the preferred approach for AFSK systems \cite{peraltaUmaAbordagemEm2018,addaimDesignLowcostTelecommunications2010}. Although the advantage of simplicity is evident, this solution performs poorly in terms of bit error rate (BER) compared to the coherent detection \cite{pacelliAlldigitalAFSKModem2020}. It directly impacts the data transmission rate of the telecommand link, which is limited to 1200 bps for most products available on the market \cite{muriSurveyCommunicationSubsystems2012}. In addition to the bit rate improvement, an increased BER can also impact the antenna design on board the satellite, where space, power, and weight limitations are demanding. However, developing a coherent demodulator imposes a higher complexity on the receiver design. The computational complexity of the symbol detection is usually the major issue as this task might be overwhelming for modulations with memory. For instance, the Viterbi algorithm, an efficient implementation of the maximum likelihood sequence detection, requires \(\order{pq}\) operations per trellis stage, where \(p\) is the number of nodes, \(q\) is the number of transitions to the next state, and \(\order{\cdot}\) denotes the big O notation \cite{moonMathematicalMethodsAlgorithms2000}.

Another concern in adopting coherent demodulation is dealing with synchronization impairments. Typically, coherent detection of continuous-phase modulation (CPM) signals is avoided because the synchronization circuits are usually complex \cite{mengaliSynchronizationTechniquesDigital1997}. Furthermore, most of the available works address this problem using analog systems, whereas, for the application at hand, due to the adopted electronic technology, a demodulation architecture with discrete-time processing is necessary. Thus, the design of modern digital modems is of paramount importance for further development and efficient operation of micro\footnote{From \(10\) kg up to \(100\) kg.}, nano\footnote{From \(1\) kg up to \(10\) kg.}, and pico\footnote{From \(0.1\) kg up to \(1\) kg.} satellite systems. The \textcolor{black}{recent} improvement of digital signal processing techniques, field-programmable gate arrays (FPGAs), and microelectronics has motivated a new generation of TT\&C transceivers that brings hardware flexibility provided by the radio-defined software concept, but, at the same time, goes along with the current modulation schemes.

\textcolor{black}{
    The recent literature has shown an ongoing interest in the FSK modulation for cutting-edge applications, such as internet of things (IoT). Although the mathematical treatment of such references is suitable for digital signal processors, they mainly focus on channel modeling for IoT environments, channel coding techniques, and theoretical error probability of coherent or noncoherent FSK modems in such environments \cite{Chandra2015BERRSBFSKIoT,Xiang2021DataEnhancementIoT,Rugini2019nonorthogonalFSKIoT}. In contrast, the present work focuses on bit detection from the received baseband signal and principally parameter estimation for carrier and timing synchronization. Although some recent articles also addressed the synchronization problem in FSK demodulators, they still tend to adopt an IoT-centric perspective, often relying on preambles defined by standards like IEEE 802.15.4, which are only suitable for IoT modems \cite{Lim2024SDRIoT,Morikawa2024HetIoT}. For satellite systems, a proper treatment of timing and carrier synchronization for FSK is still open. Moreover, the AFSK, a very specific type of FSK modulation, is poorly addressed in the literature, despite being a standard scheme for satellite communication. To the best of the authors' knowledge, there is no article focused on coherent AFSK demodulators suitable for CubeSat dynamics. The few articles that discuss AFSK modems typically adopt the noncoherent approach and do not focus on the signal processing aspects of the demodulator \cite{ahmadDesignTerminalNode2016,davisDesignSmallSatellite2015,riceBitErrorRate1981,velascoSoftwareBasedAFSK2018}.
}

In this paper, a new all-digital coherent AFSK demodulator with joint phase and timing estimator is presented. We assume that the AFSK signal passes through an additive white Gaussian noise (AWGN) channel with phase and timing impairments due to the satellite-station line-of-sight (LOS) dynamics. The Viterbi algorithm \textcolor{black}{(VA)} is used for bit detection, and a Kalman filter (KF) algorithm is used to estimate the LOS phase shift, which takes into account its first- and second-order derivative to ensure robust tracking performance under realistic CubeSat communication link scenarios. Additionally, the LOS timing delay is directly derived from the LOS phase, and the new samples are generated through interpolation. The KF estimates operate jointly and in decision-directed (DD) mode with the \textcolor{black}{VA}, i.e., the detected bit statistics are fed back to the KF to update its estimates. The system BER is analyzed through Monte Carlo simulations for different realistic scenarios of LOS dynamics, simulated using the Simplified General Perturbations 4 (SGP4) \cite{SGP4OrbitDetermination} algorithm that precisely propagates satellite orbits by using a two-line element (TLE) \cite{RevisitingSpaceTrackReport3} file. In addition, the performance of the proposed scheme is also compared to the theoretical bit error probability of the noncoherent approach (usually adopted in the literature), which highlights the main advantage of the coherent demodulator. The quality of the KF estimates are assessed using a moving root mean squared error (MRMSE) statistic, which captures how accurate the estimates are at each part of the LOS dynamic scenario, and the trace of the KF state error covariance matrix estimate, as these metrics effectively capture the filter's estimation accuracy and the overall uncertainty in the state estimates, respectively. 

The main contributions of this work are:
\begin{itemize}
    \item New decision-directed closed-loop all-digital AFSK demodulator comprised of the \textcolor{black}{VA} for bit detection and a KF for synchronization\textcolor{black}{;}
    \item \textcolor{black}{Original mathematical derivation of a joint synchronization and demodulation algorithm based on the VA and a KF;}
    \item Realistic simulation and performance evaluation of the proposed joint synchronization and demodulation scheme under different scenarios of CubeSat LOS dynamics orbiting in LEO\textcolor{black}{;}
\end{itemize}

This paper is organized as follows: the AFSK signal and the channel model are presented in Section \ref{sec:signal} and \ref{sec:channel}, respectively. In Section \ref{sec:demodulator}, the AFSK demodulator is detailed. Section \ref{sec:results} is devoted to evaluating the proposed demodulator under realistic LOS dynamics scenarios. Finally, in Section \ref{sec:conclusions}, our conclusions are outlined.

\subsection{Notation}

Continuous- and discrete-time signals are indexed in parenthesis and brackets, respectively. For instance, \(s \left( t \right)\) and \(s \left[ n \right]\) represent a continuous-time signal and its discrete-time version at a given sampling time. For the continuous-time case, superscripts dots are used to denote derivatives. For example, \(\dot{s} \left( t \right)\) and \(\ddot{s} \left( t \right)\) denote the first- and second-order derivative with respect to \(t\). The matrix \(\mathbf{0}_{p \times q}\) indicates a zero matrix with \(p\) rows and \(q\) columns. The positive underscore superscript in \(\mathbb{Z}^+\) or \(\mathbb{R}^+\) denotes the set of nonnegative integers or real numbers, respectively. The notations \(a(t) \sim \mathcal{N}(\mu_a, \sigma_{a}^{2})\) and \(b(t) \sim \mathcal{CN}(\mu_b, \sigma_{b}^{2})\) denote a real- and complex-valued Gaussian process with mean \(\mu_a\) and \(\mu_b\), and variance \(\sigma_{a}^{2}\) and \(\sigma_{b}^{2}\), respectively. Moreover, the notation \(a \left[ n \right] \sim \mathcal{U}\left( a,b \right)\) denotes a uniformly distributed stochastic process within the range \(\left[ a, b \right]\).

%% file: secs/channel.tex
\section{AFSK Signal} \label{sec:signal}

The AFSK is a binary modulation that uses audible frequencies to transmit the data. The switching from one frequency to another can be accomplished using two different oscillators tuned to the desired frequencies. Alternatively, the AFSK modulator can preserve the continuity of the phase by generating a continuous-phase frequency-shift keying (CPFSK) signal. In this case, instead of switching the signals from two different oscillators, a unique oscillator with constant frequency is used. However, its instantaneous phase varies linearly depending on the transmitted bit. This results in a continuous-phase signal whose instantaneous frequency switches between the desired frequencies. Considering the continuous-phase implementation, the pre-envelope of the transmitted AFSK signal can be defined as
\begin{align}
  \label{eq:s_RF_t}
  s_{RF}\left(t\right) & = \sqrt{\frac{2E_b}{T}} \text{cos}{\big(2\pi f_{RF}t + \Phi \left(t, \mathcal{A}\right) \!\big)}\text{,}
\end{align}
where \(E_b\) is the bit energy, \(f_{RF}\) is the carrier radio frequency, \(t\) is the time instant, \(\Phi \left(t, \mathcal{A}\right) \in \mathbb{R}\) is the time-varying phase of the AFSK signal, which depends on the transmitted symbol sequence \(\mathcal{A}=\left\{\alpha \left[ 0 \right],\alpha\left[ 1 \right],\dots,\alpha\left[ k \right]\right\} \ \forall \ k \in \mathbb{Z}^+\) with \(\alpha \left[ k \right]\in \left\{-1,1\right\}\) being the \(k\)th symbol with duration of \(T\) seconds. The signal \(\Phi \left(t, \mathcal{A}\right)\) can be expressed in the interval \(k T \leq t < \left(k + 1\right)T\) as 
\begin{align}
  \label{eq:Phi_talpha}
  \Phi \left(t, \mathcal{A} \right) & = 2\pi h \sum_{i=0}^k \alpha \left[ i \right] q\left(t-iT\right) \nonumber \\ 
                                 & = 2\pi h \alpha \left[ k \right] q\left(t-k T\right) + 2\pi h \sum_{i=0}^{k -1} \alpha \left[ i \right] q\left(t-iT\right) \nonumber \\ 
                                 & = 2\pi h \alpha \left[ k \right] q\left(t-k T\right) + \pi h \sum_{i=0}^{k -1} \alpha \left[ i \right] \nonumber \\ 
                                 & = \theta \left[ k \right] + \theta\left(t,\alpha\left[ k \right]\right)\text{,} 
\end{align}
where 
\begin{align}
  q\left( t\right) & = \left\{ \begin{array}{rl}
    0 & \text{for } t<0, \\[2mm]
    \displaystyle{ \frac{t}{2T} } & \text{for } 0\le t \le T  \\[2mm]
    \displaystyle{\frac{1}{2}} & \text{for } t > T \\[2mm]
\end{array} \right. 
\end{align}
is the phase pulse, given by \cite{proakisDigitalCommunications2007}, \(\theta\left(t,\alpha \left[ k \right]\right) = 2\pi h \alpha \left[ k \right] q\left(t-k T\right)\) is the transition phase during \(k T \leq t < \left(k + 1\right)T\), and
\begin{align}
  \label{eq:theta_k}
  \theta \left[ k \right]  & = \pi h \sum_{i=0}^{k -1} \alpha \left[ i \right]
\end{align}
is the phase memory up to the symbol \(\alpha \left[ k-1 \right]\). The variable \(h \in \mathbb{R}\) indicates the modulation index and defines how much the time-varying phase increases or decreases for each transmitted symbol. If \(h\) is rational, when \(t = kT\), \(\Phi \left(t, \mathcal{A} \right) = \theta \left[ k \right] \in \mathcal{B}\), where \(\mathcal{B}\) is a finite set of terminal phases. The cardinality of \(\mathcal{B}\) depends on \(h\) since each terminal phase is defined in intervals of \(\pm \pi h\).  The modulation index is given by
\begin{align}
  h & = T\Delta f,
\end{align}
where \(\Delta f\) is the difference between the frequencies used to transmit the data.

Previous works have reported the use of the AFSK signalling with a bit rate of \(R=1/T=1200\) bps (bits per second) and with a frequency difference of \(\Delta f = 1\) kHz \cite{peraltaUmaAbordagemEm2018,addaimDesignLowcostTelecommunications2010}. Under these conditions, \(h = 5/6\) and its terminal phases are given by \cite{pacelliAlldigitalAFSKModem2020}
\begin{align}
  \mathcal{B} & = \left\{ \beta_0 ,\beta_1 ,\ldotp \ldotp \ldotp ,\beta_{11} \right\} \nonumber \\ 
  & = \left\{ 0,\frac{5\pi }{6},\frac{10\pi }{6},\ldotp \ldotp \ldotp ,\frac{55\pi }{6}\right\} \text{.}
  \label{eq:set_B}
\end{align}
Therefore, the AFSK signalling contains \(\#\mathcal{B} = 12\) terminal phases, where \(\#\cdot\) denotes the cardinality operation. The phase \(\Phi \left(t, \mathcal{A} \right)\) decreases or increases \(5\pi/6\) rad depending upon whether the $k$th transmitted bit is \textcolor{black}{$\alpha \left[ k \right]=-1$ or $\alpha \left[ k \right]=1$}, respectively. \textcolor{black}{This creates terminal phases that are multiples of \(\pi/6\) around the unit circle}.  In case the terminal phase would fall outside the extremes $\beta_0$ and $\beta_{11}$, it wraps around these values to stay inside the set $\mathcal{B}$.

\section{Channel Model}
\label{sec:channel}

The uplink frequency lies within the range of 148-150 MHz \cite[Chapter 5]{sebestyenLowEarthOrbit2018}. At these frequencies, tropospheric effects can be neglected because the troposphere is nondispersive only for frequencies up to 15 GHz \cite[p. 122]{subiranaGNSSDataProcessing}. Consequently, group and phase delays remain nearly constant\footnote{The tropospheric delay can be considered constant for the same obliquity factor, provided that the wet and hydrostatic components of the tropospheric medium do not vary significantly \cite[Chapter 5]{subiranaGNSSDataProcessing}.}, and their impact can be accurately modeled and removed by using suitable tropospheric models. Additionally, this work assumes negligible refractive ionospheric effects as CubeSats in low Earth orbit (LEO) typically remain in the ionosphere's lower layers, where this disturbance is mild \cite{SpaceWeatherOnLEOOrbits}. Consequently, the phase and group delays for the uplink carrier frequency are nearly constant and can be effectively corrected by using ionospheric models \cite{vasylyevModelingIonosphericScintillation2022}.

Given these considerations, let us now consider the channel effects caused by satellite-station line-of-sight (LOS) dynamics. Specifically, the channel introduces a delay \(\tau_{LOS} \left( t \right)\), and the transmitted signal \(s_{RF} \left( t \right)\) is corrupted by additive white Gaussian noise (AWGN). Therefore, the received signal can be modeled as
\begin{align}
  \label{eq:r_RF_t_1}
  r_{RF} \left(t\right) & = A(t)s_{RF}\big(t-\tau_{LOS} \left( t \right) \! \big) + \eta_{RF}\left(t\right),
\end{align}
where \(\eta_{RF}\left( t \right)\) is a real, zero-mean, and white Gaussian noise with two-sided power spectrum density equals to \(N_0/2 \in \mathbb{R}^+\), and \(A \left( t \right) \in \mathbb{R}\) models the received signal amplitude variations due to the antenna gain, free-space path loss, and other attenuation factors. By substituting \eqref{eq:s_RF_t} into \eqref{eq:r_RF_t_1}, we have
\begin{align}
    r_{RF} \left(t\right) & = A(t)\sqrt{\frac{2E}{T}} \text{cos}\bigg(2\pi f_{RF}\big(t-\tau_{LOS} \left( t \right) \! \big) + \Phi \Big(t-\tau_{LOS} \left( t \right), \mathcal{A} \Big) \bigg) +  \eta_{RF}\left(t\right).
\end{align}
By observing the LOS phase evolution from an initial time instant \(t_0\), it follows that
\begin{align}
    r_{RF} \left(t\right) &= A(t)\sqrt{\frac{2E}{T}} \text{cos}\Bigg(2\pi f_{RF}\bigg(t-\tau_{LOS}(t_0) - \int^{t}_{t_0}{\frac{\partial\tau_{LOS}(\color{black}u)}{\partial \color{black}u} \color{black} \dd{u}} \bigg) \nonumber\\
  &\quad + \Phi \Big(t-\tau_{LOS} \left( t \right), \mathcal{A} \Big) \Bigg) +  \eta_{RF}\left(t\right) \nonumber \\
  &= A(t)\sqrt{\frac{2E}{T}} \text{cos}\Bigg(2\pi f_{RF}t + \phi_{LOS}(t_0) + 2\pi\int^{t}_{t_  0}{f_D(\color{black}u) \color{black} \dd{u}} + \nonumber \\
  &\quad + \Phi \Big(t-\tau_{LOS} \left( t \right), \mathcal{A} \Big) \Bigg) +  \eta_{RF}\left(t\right),
\end{align}
where
\begin{align}
    f_D\left( t \right) = - f_{RF}\frac{\partial\tau_{LOS}(t)}{\partial t}
\end{align}
is the Doppler frequency shift and \(\phi_{LOS}(t_0) = - 2\pi f_{RF}\tau_{LOS}(t_0)\). Finally,
\begin{align}
    r_{RF} \left(t\right) &= A(t)\sqrt{\frac{2E}{T}} \text{cos}\bigg(2\pi f_{RF}t + \phi_{LOS}(t) + \Phi \Big(t-\tau_{LOS} \left( t \right), \mathcal{A} \Big) \bigg) + \eta_{RF}\left(t\right) \text{,}
\end{align}
where
\begin{align}
  \phi_{LOS}(t) = \phi_{LOS}(t_0) + 2\pi\int^{t}_{t_0}{f_D(\color{black}u) \color{black} \dd{u}}.
\end{align}

\begin{figure}[H]
  \centering
  \includegraphics{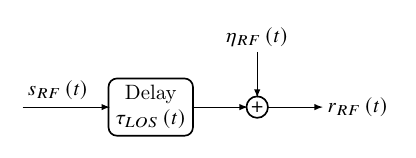}
  \caption{Channel model.}
  \label{fig:channel}
\end{figure}

Figure \ref{fig:channel} shows a complete block diagram illustrating the assumed channel model.

%% file: secs/demodulator.tex
\section{AFSK Demodulator} 
\label{sec:demodulator}

The transmitted signal is received by a set of antennas and delivered to the RF front-end module, where an analogue preprocessing step is performed. The received signal is shifted to an intermediate frequency and then passes through an automatic gain control (AGC) that adjusts its amplitude to the full-scale level of the analog-to-digital converter (ADC). This signal is then sampled by the ADC at a rate of \(f_s = 1/T_s\) and subsequently downconverted to baseband in digital means, yielding
\begin{align}
  \label{eq:r_tilde_n}
  \tilde{r} \left[ n \right] = \tilde{s}\left[ n \right]e^{j\phi_{LOS}\left[ n \right]} + \tilde{\eta}\left[ n \right] \text{,}
\end{align}
where 
\begin{align}
  \label{eq:discrete_s_n}
  \tilde{s}\left[ n \right] = s\big(nT_s - \tau_{LOS}\left[ n \right]\big) \in \mathbb{C}
\end{align}
characterizes the discrete-time version of the transmitted AFSK complex envelope. Here, \(n \in \mathbb{N}^+\) represents the sampling instants and the tilde notation indicates that the LOS effects have not been corrected yet. In addition, \(\tau_{LOS}\left[ n \right]\) and \(\phi_{LOS}\left[ n \right]\) model the discrete-time delay in seconds and phase shift in radians, respectively. The distortions caused by each downcoversion step, as well as the quantization, are assumed to be negligible. Therefore, it is also justified to assume that the remaining additive noise of the received signal \(\tilde{\eta}\left[ n \right] \sim \mathcal{CN}\left( 0, 2N_0B/T_s \right)\) is a zero-mean complex Gaussian noise with power spectral density of \(N_0/T_s\) within \(\left[ -B, B \right]\) Hz, i.e., the bandwidth where the baseband signal is allocated.

The proposed AFSK demodulator is divided in two main parts: The \textcolor{black}{VA} which performs the maximum likelihood sequence detection (MLSD) for the bits, and a synchronization system based on a KF that is able to generate precise estimates of the LOS dynamics. Both parts interact mutually in a DD manner, thus providing significantly enhanced performance if compared with noncoherent bit detection, as shown in Section \ref{sec:results}.

\subsection{Line-of-sight dynamics correction}
\label{subsec:LOS_correction}
In order to perform coherent bit detection, it is necessary to first correct the timing delay and phase shift caused by the LOS dynamics on the received signal given by \eqref{eq:r_tilde_n}. Suppose momentarily that estimates of \(\phi_{LOS} \left[ n \right]\) and \(\tau_{LOS}\left[ n \right]\) are available at the sampling time instant \(n\). The KF algorithm, that provides such estimates, is detailed in Section \ref{sec:KF_section}.

\textcolor{black}{It is important to anticipate here that, although the timing delay and carrier phase estimates provided by the KF are well-suited for CubeSat applications, they are not perfect. The impacts of large errors in these estimates on the performance of the VA will be carefully analyzed in Section \ref{subsec:Baysian_carrier_phase_tracking_system}.}

Firstly, the LOS phase shift is corrected by mixing the received signal with samples of a carrier replica, i.e., a complex exponential using the estimated carrier phase, \(\hat{\phi}_{LOS} \left[ n \right]\). Here, the hat notation refers to estimates. Secondly, the timing delay correction is performed. However, an all-digital demodulator cannot directly act at the sampling instant of the ADC as it operates in free-running mode. Therefore, it is necessary to use estimates of the timing delay \(\hat{\tau}_{LOS}\left[ n \right]\) to generate the interpolants, \(r\left[ m \right]\). Note that the time index changed here to explicitly denote the interpolation step. Nevertheless, both time indices have the same sampling time. There are many interpolator architectures that could be used to perform this task. In this case, we assume that an interpolator with a Farrow architecture is used \cite{farrowContinuouslyVariableDigital1992}.

Thus, the received signal given by \eqref{eq:r_tilde_n}, after the phase shift and the timing delay corrections, can be written as
\begin{align}
  \label{eq:r_corrected}
  r \left[ m \right] =
  s\left[ m \right]e^{j\varepsilon_{\phi}\left[ m \right]} + \eta\left[ m \right] \text{,}
\end{align}
where
\begin{align}
  \label{eq:s_m}
  s\left[ m \right] &= s\big(mT_s\big) = e^{j\Phi\left[ m, \mathcal{A}\right] } 
\end{align}
is the corrected discrete-time complex envelope, with
\begin{align}
  \label{eq:discrete_Phi}
  \Phi\left[m, \mathcal{A} \right] = \theta\left[ k \right] + \theta\left[m-kN,\alpha\left[k\right]\right]
\end{align}
being the sampled version of \(\Phi\left(t, \mathcal{A} \right)\),
\begin{align}
  \label{eq:phase_error}
  \varepsilon_{\phi}\left[ m \right] = \phi_{LOS}\big(mT_s + \hat{\tau}_{LOS}\left[ m \right] \big) - \hat{ \phi}_{LOS}\big(mT_s + \hat{\tau}_{LOS}\left[ m \right] \big)
\end{align}
denotes the carrier phase residual, and
\begin{align}
  \label{eq:w_m}
  \eta\left[ m \right] = \tilde{\eta}\big(mT_s + \hat{\tau}_{LOS}\left[ m \right] \big) e^{-j\hat{ \phi}_{LOS}\big(mT_s + \hat{\tau}_{LOS}\left[ m \right] \big)}
\end{align}
represents a Gaussian noise with its power spectral density shifted. It is important to note that even though the noise has been shifted, its first- and second-order moments do not change as the stochastic process is assumed to be weak-sense stationary \cite{papoulisProbabilityRandomVariables2001}.

\subsection{The Viterbi \textcolor{black}{algorithm (VA)}}
The LOS-corrected signal, given by \eqref{eq:r_corrected}, passes through a filter bank composed by a group of matched filters that correlates it with the possible transmitted complex envelopes. Such correlation is used as a metric by the \textcolor{black}{VA} to perform the maximum likelihood sequence detection (MLSD). The algorithm searches through the trellis for the sequence whose correlation is maximum \cite{proakisDigitalCommunications2007,sklarHowLearnedLove2003}. Therefore, let us define
\begin{align}
  g \left[m\right] &= r\left[m\right] \ast h_{MF}\left[m\right] \nonumber \\
  =& \sum_{i=m-N+1}^{m} r\left[i\right] e^{-j\Phi \left[i+N-1-m, \mathcal{A}\right]}.
  \label{eq:g_m1}
\end{align}
as one of the outputs of the filter bank, where
\begin{align}
  h_{MF} \left[m\right] = & \left\{ \begin{array}{lr}
    e^{- j\Phi \left[N-1-m, \mathcal{A}\right]}& \text{for } 0 \leq m \leq N-1 \\
    0 & \text{otherwise}
  \end{array} \right.
  \label{eq:h1_m}
\end{align}
is its impulse response and \(\ast\) is the convolution operator.

At the demodulator, we use the notation \(\hat{\alpha}\left[k\right]\) to indicate that it is the detected symbol, which may or may not differ from the transmitted one, \(\alpha\left[k\right]\). Therefore,
\begin{align}
  \label{eq:hat_theta_k}
  \hat{\theta}\left[k\right]  & = \pi h \sum_{i=0}^{k-1} \hat{\alpha}\left[i\right] \in \mathcal{B}
\end{align}
denotes the estimated terminal phase of the transmitted signal, i.e, the phase memory.

It is possible to notice that decimating at \(N(k+1)-1\) and taking the real part of \eqref{eq:g_m1} yields the correlation metric between the \(k\)th transmitted and received symbol \cite[Section 9.2]{mengaliSynchronizationTechniquesDigital1997}. In other words, the real part of the statistic
\begin{align}
  \label{eq:v_kk}
  g \left(\hat{\alpha}\left[k\right],\hat{\theta}\left[k\right]\right) = g \left[m\right]\bigg\rvert_{m = N(k+1)-1} = \sum_{i=k N}^{N(k+1)-1} r\left[i\right] e^{-j\Phi \left[i-kN, \mathcal{A}\right]}
\end{align}
corresponds to the \(k\)th transition metric for the symbol \(\hat{\alpha}\left[k\right]\), during the interval \(k N \leq m \leq N(k+1)-1\). We can rewrite \eqref{eq:v_kk} as
\begin{align}
  \label{eq:g_k_k}
  g \left(\hat{\alpha}\left[k\right],\hat{\theta}\left[k\right]\right)&= \sum_{i=k N}^{N(k+1)-1} r\left[i\right] e^{-j\Phi \left[i-kN, \mathcal{A}\right]} \nonumber \\
  &= \sum_{i=k N}^{N(k+1)-1} r\left[i\right] e^{-j\left(\hat{\theta}\left[k\right] + \theta\left[i-kN,\hat{\alpha}\left[k\right]\right]\right)} \nonumber \\
  &=  e^{-j\hat{\theta}\left[k\right]} \sum_{i=k N}^{N(k+1)-1} r\left[i\right] e^{-j\theta\left[i-kN,\hat{\alpha}\left[k\right]\right]} \nonumber \\
  &=  e^{-j\hat{\theta}\left[k\right]} c \left(\hat{\alpha}\left[k\right]\right).
\end{align}
The first term of \eqref{eq:g_k_k}, \(e^{-j\hat{\theta}\left[k\right]}\), refers to the phase memory of the AFSK signal, which depends on the entire sequence of the previously detected symbols (\textit{cf.} \eqref{eq:hat_theta_k}). The second term of \eqref{eq:g_k_k},
\begin{align}
    \label{eq:c_k_k}
    c \left(\hat{\alpha}\left[k\right]\right) = c \left[ m \right]\bigg\rvert_{m = N(k+1)-1} = \sum_{i=k N}^{N(k+1)-1} r\left[i\right] e^{-j\theta\left[i-kN,\hat{\alpha}\left[k\right]\right]},
\end{align}
is obtained after passing \(r\left[m\right]\) through a filter bank whose impulse responses depend on the possible transmitted symbol \(\hat{\alpha}\left[ k \right] \in \left\{ -1, 1 \right\}\) and are given by
\begin{IEEEeqnarray}{rCl}
  \label{eq:h}
  h_{FB} \left[m\right] & = & \left\{ \begin{array}{lr}
      e^{- j\theta\left[N-1-m,\hat{\alpha}\left[k\right]\right]} & \text{if } 0 \leq m \leq N-1 \\
      0 & \text{otherwise}
\end{array} \right. .
\end{IEEEeqnarray}
Note that, instead of using a bank with 24 filters (\(\# \mathcal{B} = 12\) terminal states with two possible symbols each) to yield \(g \left[ m \right]\), one can use just two filters to yield \(c \left[ m \right]\), as shown in \eqref{eq:c_k_k}. The statistic \(c \left(\hat{\alpha}\left[k\right]\right)\) is generated by decimating the filter bank output at the instant \(N(k+1)-1\). After this stage, the demodulator operates at the symbol rate. Afterwards, \(c \left(\hat{\alpha}\left[k\right]\right)\) is multiplied by the phase memory inherent of the continuous phase modulation, as shown in \eqref{eq:g_k_k}. However, since \(\hat{\theta}\left[k\right] \in \mathcal{B}\) is unknown to the demodulator, a metric with all terminal phases is computed and delivered to the \textcolor{black}{VA}, which recovers the memory \(\hat{\theta}\left[k\right]\) and detects the bits. Mathematically, the metric
\begin{align}
    \label{eq:G_k}
  \mathbf{G}\left[k\right] = \mathbf{c}\left[k\right] \boldsymbol{\betaup}^\top \in \mathbb{C}^{2 \times 12}
\end{align}
is the input of the  \textcolor{black}{VA}, where
\begin{align}
  \mathbf{c}\left[k\right] \triangleq \begin{bmatrix}
    c(\hat{\alpha}\left[k\right]=-1) & c(\hat{\alpha}\left[k\right]=1)\end{bmatrix}^\top \in \mathbb{C}^{2}
\end{align}
is the output of the filter bank, and
\begin{align}
  \boldsymbol{\betaup} \triangleq& \begin{bmatrix}
    e^{-j\beta_0}& e^{-j\beta_1} & \cdots & e^{-j\beta_{11}}
   \end{bmatrix}^\top \in \mathbb{C}^{12}
\end{align}
is a vector of complex exponentials whose phase argument is an element of the set \(\mathcal{B}\) (see \eqref{eq:set_B}). The \textcolor{black}{VA} takes the real part of \(\mathbf{G}\left[k\right]\) and searches through the trellis for the path whose likelihood function is maximum (i.e., the survivor path), thus detecting the transmitted bits, \(\hat{\alpha}\left[ k \right]\) \cite{sklarHowLearnedLove2003}.

\subsection{LOS phase and timing estimation using a Kalman filter (KF)}
\label{sec:KF_section}
The problem of recursive estimation of the phase shift and timing delay caused by the LOS dynamics can be tackled by implementing classical solutions based on phase-locked loops and frequency-locked loops \cite{kaplanUnderstandingGPSGNSS2017}, or Bayesian estimators \cite{brownIntroductionRandomSignals1997}, such as the standard KF \cite{ArePLLsDeadVila} and its variants\footnote{The most prominent variants are the extended KF, the unscented KF \cite{UnscentedKalmanFilter}, and the Cubature KF \cite{CubatureKalmanFilter}.}.

The main advantage of using Bayesian estimators for synchronization are their flexibility to incorporate other impairments, such as ionospheric scintillation \cite{fohlmeisterDualKalmanFiltering2018,lopesIonosphericScintillationMitigation2023,locubiche-serraDoublyadaptiveAutoregressiveKalman2016} and multipath \cite{MultiFrequencyKalmanFilteringforJointIonospheric}, as well as to adaptively change its filter bandwidth based on the SNR of its measurements. Therefore, this class of carrier phase tracking systems is very suitable for next generation software-defined radio (SDR) receivers \cite{SurveySalcedo} and \cite{SurveyVila}. With that, a KF to operate in DD mode with the \textcolor{black}{VA} is presented hereafter.

\subsubsection{Carrier phase LOS dynamics model}
\label{sec:LOS_model}
As presented in \cite[Section 2.4]{fohlmeisterDualKalmanFiltering2018}, the relative range and velocity between the satellite and the ground station can be modeled by employing a nth-order state-space Wiener model. The input vector of this state-space model is the driving Gaussian noise of the Wiener process, denoted as \(\xi_W \left( t \right) \sim \mathcal{N} \left( 0, \sigma_W^2 \right)\). If \(w_W \left( t \right) \in \mathbb{R}\) is a Wiener process driven by \(\xi_W \left( t \right)\), then \(\dot{w}_W \left( t \right) = \xi_W \left( t \right)\) and \(w_W \left( t + \tau \right) - w_W \left( t \right) \sim \mathcal{N}\left( 0, \tau \sigma_W^2 \right)\).

The reader should note that, the higher the order of the model, the more accurate the representation of the carrier phase LOS dynamics. In this work, a third-order Wiener model is considered, which is sufficient to capture the main effects for CubeSats applications. Therefore, we have
\begin{align}
    \label{eq:Wiener_model}
  {\renewcommand{\arraystretch}{1.5}
  \underbrace{
  \begin{bmatrix}
    \dot{\phi}_{LOS}\left( t \right) \\
    \dot{f}_D\left( t \right) \\
    \ddot{f}_D\left( t \right)
  \end{bmatrix}}_{\textcolor{black}{=} \dot{\mathbf{x}}\left( t \right)}}
  &=
  \underbrace{\begin{bmatrix}
    0 & 2\pi & 0 \\
    0 & 0 & 1 \\
    0 & 0 & 0
  \end{bmatrix}}_{\textcolor{black}{=}\mathbf{F}_W}
  \underbrace{\begin{bmatrix}
    \phi_{LOS}\left( t \right) \\
    f_D\left( t \right) \\
    \dot{f}_D\left( t \right)
  \end{bmatrix}}_{\textcolor{black}{=}\mathbf{x}\left( t \right)}
  +
  \underbrace{\begin{bmatrix}
    \xi_\phi\left( t \right) \\
    \xi_{f_D}\left( t \right) \\
    \xi_{\dot{f}_D}\left( t \right)
  \end{bmatrix}}_{\textcolor{black}{=} \bm{\xi}\left( t \right)} \nonumber \\
  &= \mathbf{F}_W \mathbf{x}\left( t \right) + \bm{\xi}\left( t \right) \text{,}
\end{align}
where \(\mathbf{x}(t) = \big[\phi_{LOS}(t),f_D\left( t \right),\dot{f}_D\left( t \right)\big]^\top\) is the continuous-time state variable, \(\mathbf{F}_W\) is known as the state transition matrix, and the vector \(\bm{\xi}\left( t \right) = \big[\xi_\phi\left( t \right),\xi_{f_D}\left( t \right),\xi_{\dot{f}_D}\left( t \right)\big]^\top \sim\mathcal{N}\big(0,\mathbf{Q_{\bm{\xi}}}\big)\) is a continuous-time multivariate white Gaussian noise that captures all uncertainties in the relative movement between the ground station and the satellite as well as the phase drift caused by clock aging. Therefore, in a perfect scenario without uncertainties, \(\bm{\xi}\left( t \right)\) would be equal to \(\mathbf{0}_{3\times1}\) and the carrier LOS phase evolution would be given by \(\mathbf{F}_W \mathbf{x}\left( t \right)\).

Thus, the cross-covariance matrix of \(\bm{\xi}\left( t \right)\) for two different time instants \(t_1\) and \(t_2\) can be written as
\begin{align}
  \label{eq:continuous_covariance}
  \mathbb{E}\big\{\bm{\xi}\left( t_1 \right)\bm{\xi}^\top\left( t_2 \right)\big\} = \begin{cases}
    \mathbf{Q}_{\bm{\xi}} = \text{diag} \left( \sigma_1^2, \sigma_2^2, \sigma_3^2 \right) & \text{if } t_1 = t_2 \\
    \mathbf{0}_{3 \times 3} & \text{otherwise}
  \end{cases} \text{,}
\end{align}
where \(\mathbb{E}\{\cdot\}\) and \(\text{diag} \left( \cdot \right)\) is the expectation and diagonalization operator, and \(\sigma_1^2\), \(\sigma_2^2\), and \(\sigma_3^2\) denote the variance of \(\xi_\phi\left( t \right)\), \(\xi_{f_D}\left( t \right)\), and \(\xi_{\dot{f}_D}\left( t \right)\), respectively.

The solution for \eqref{eq:Wiener_model}, according to \cite[Appendix A2]{fohlmeisterDualKalmanFiltering2018}, reads
\begin{align}
  \label{eq:Wiener_solution}
  \mathbf{x} \left( t \right) = e^{\left( t-t_0 \right) \mathbf{F}_W}\mathbf{x}\left( t_0 \right) + \int^{t}_{t_0}{e^{\left( t-\tau \right) \mathbf{F}_W}\bm{\xi}\left( \tau \right)\dd{\tau}} \text{.}
\end{align}
By using the Taylor series expansion, \(e^{\left( t-\tau \right) \mathbf{F}_W}\) can be rewritten as follows
\begin{align}
  \mathbf{A}\left( t-\tau \right) \triangleq e^{\left( t-\tau \right) \mathbf{F}_W} = \sum_{i=0}^{\infty}{\frac{\big( \left( t-\tau \right) \mathbf{F}_W \!\big)^i}{i!}} \text{.}
\end{align}
Therefore, noticing that all elements of this sum for \(i>2\) are equal to \(\mathbf{0}_{3 \times 3}\), it is possible to write
\begin{align}
  \label{eq:Taylor_expansion}
  \mathbf{A}\left( t-\tau \right) &= 
  \begin{bmatrix}
    1&0&0\\
    0&1&0\\
    0&0&1
  \end{bmatrix}
  +
  \begin{bmatrix}
    0&2\pi\left( t-\tau \right)&0\\
    0&0&\left( t-\tau \right)\\
    0&0&0
  \end{bmatrix}
  +
  \begin{bmatrix}
    0&0&\pi \left( t-\tau \right)^2 \\
    0&0&0 \\
    0&0&0
  \end{bmatrix} \notag\\
  & =
  \begin{bmatrix}
    1&2 \pi \left( t-\tau \right)& \pi \left( t-\tau \right)^2\\
    0&1& \left( t-\tau \right)\\
    0&0&1
  \end{bmatrix} \text{.}
\end{align}
Furthermore, by substituting \eqref{eq:Taylor_expansion} into the first and second terms on the right side of \eqref{eq:Wiener_solution}, and considering time discretization such that \(t=kT\) and \(t_0=\left( k-1 \right)T\), we get
\begin{align}
    \label{eq:discrete_Wiener_solution}
  \mathbf{x}\left[ k \right]&= 
  \underbrace{\begin{bmatrix}
    1&2 \pi T& \pi T^2\\
    0&1& T\\
    0&0&1
  \end{bmatrix}}_{\textcolor{black}{=}\mathbf{F}} \mathbf{x}\left[ k-1 \right]
   +\underbrace{\int^{kT}_{\left( k-1 \right)T}{
    \mathbf{A}(kT-\tau) \bm{\xi}\left( \tau \right)\dd{\tau}
    }}_{\textcolor{black}{=}\mathbf{w}[k]} \notag\\
  &= \mathbf{F}\mathbf{x}\left[ k-1 \right] + \mathbf{w}[k] \text{,}
\end{align}
where \(\mathbf{F}\) is the discrete-time state transition matrix and \(\mathbf{w}\left[ k \right] \sim \mathcal{N}\big(0,\mathbf{Q}\big)\) is a multivariate Gaussian noise whose covariance matrix can be calculated as
\begin{align}
  \label{eq:discrete_cross_covariance}
  \mathbf{Q} & = \mathbb{E}\big\{ \mathbf{w}[k]\mathbf{w}^\top[k] \big\} \notag \\
             & = \mathbb{E}\bigg\{ \int^{kT}_{\left( k-1 \right)T}{
                \mathbf{A}(kT-\tau_1)\bm{\xi}\left( kT - \tau_1 \right)\dd{\tau_1}
                }
                \int^{kT}_{\left( k-1 \right)T}{
                \bm{\xi}^\top\left( kT - \tau_2 \right)\mathbf{A}^\top(kT-\tau_2)\dd{\tau_2}
                } \bigg\} \notag \\
    & = \iint^{kT}_{\left( k-1 \right)T}{
      \mathbf{A}(kT-\tau_1)
      \mathbb{E}\big\{ \bm{\xi}\left( kT - \tau_1 \right)\bm{\xi}^\top\left( kT - \tau_2 \right) \big\}
      \mathbf{A}^\top(kT-\tau_2)
      \dd{\tau_1} \dd{\tau_2}}
      \text{.}
\end{align}
Now, recalling \eqref{eq:continuous_covariance} and that the values of the continuous-time covariance matrix are time-invariant, one can assume that \(k = 1\) in \eqref{eq:discrete_cross_covariance} without loss of generality. Hence, we have
\begin{align}
  \label{eq:double_int_Q}
  \mathbf{Q} =
  \iint^{T}_{0}{
      \mathbf{A}(T-\tau_1)
      \mathbf{Q}_{\bm{\xi}}
      \delta\left( \tau_1 - \tau_2 \right)
      \mathbf{A}^\top(T-\tau_2)
      \dd{\tau_1} \dd{\tau_2}
  } \text{,}
\end{align}
where \(\delta\left( t \right)\) is the Dirac delta function.

By using the iterated integral technique \cite[Section 15.2]{stewartMultivariableCalculus7th2011} and the properties of the Dirac delta function for \(\tau_1 = \tau_2 = \tau\), the equation \eqref{eq:double_int_Q} can be rewritten as
\begin{align}
    \label{eq:w_Q}
  \mathbf{Q} = 
  \int^{T}_{0}{
      \mathbf{A}\left( T-\tau \right)
      \mathbf{Q}_{\bm{\xi}}
      \mathbf{A}^\top(T-\tau)
      \dd{\tau}} \text{.}
\end{align}
Note that \(\mathbf{w}\left[ k \right] \sim \mathcal{N}\big(0,\mathbf{Q}\big)\) is the discrete-time counterpart of \(\bm{\xi}\left( t \right) \sim \mathcal{N}\big(0,\mathbf{Q_{\bm{\xi}}}\big)\) (\textit{cf.} equations \eqref{eq:discrete_Wiener_solution} and \eqref{eq:Wiener_model}). That is, \(\mathbf{w}\left[ k \right]\) models the uncertainties that disturb the carrier LOS phase measurements in the discrete-time state-space Wiener model, where its covariance matrix is given in terms of the covariance of \(\bm{\xi}\left( t \right)\) as shown in \eqref{eq:w_Q}. \textcolor{black}{By e}valuating this integral using MATLAB's symbolic math package, it is possible to obtain:
\begin{align}
  \label{eq:state_covariance}
  \mathbf{Q} =
  \sigma^2_1
  \begin{bmatrix}
    T & 0 & 0 \\
    0 & 0 & 0 \\
    0 & 0 & 0
  \end{bmatrix}
  +\sigma^2_2
  \begin{bmatrix}
    \frac{4 T^3 \pi^2}{3} & \pi T^2 & 0 \\
    \pi T^2 & T & 0 \\
    0 & 0 & 0
  \end{bmatrix}+\sigma^2_3
  \begin{bmatrix}
    \frac{\pi^2 T^5}{5} & \frac{\pi T^4}{4} & \frac{\pi T^3}{3} \\
    \frac{\pi T^4}{4} & \frac{T^3}{3} & \frac{T^2}{2} \\
    \frac{\pi T^3}{3} & \frac{T^2}{2} & T
  \end{bmatrix} \text{.}
\end{align}
The covariance matrix elements presented here are slightly different if compared with the results shown in \cite[Equation 2.60]{fohlmeisterDualKalmanFiltering2018}, since the phase in this work is modeled in radians, instead of cycles. Hence, from \eqref{eq:state_covariance}, it is easy to notice that characterizing the uncertainties that disturbs the carrier LOS phase measurements means defining the variances \(\sigma_1^2\), \(\sigma_2^2\) and \(\sigma_3^2\).

\subsubsection{Measurement model}
\label{sec:measurement_model}
Assuming correct symbol detection provided by the \textcolor{black}{VA}, i.e., \(\hat{\alpha}\left[ k \right] = \alpha\left[ k \right]\), the residual phase of the filtered signal, as given by \eqref{eq:g_k_k}, can be used as an input to an estimator to perform carrier phase recovery. In order to illustrate this, we may further elaborate \eqref{eq:g_k_k}, recalling  \eqref{eq:r_corrected}, \eqref{eq:s_m}, \eqref{eq:discrete_Phi}, and \eqref{eq:w_m} as
\begin{align}
  \label{eq:g_open}
  &g \left(\hat{\alpha}\left[k\right],\hat{\theta}\left[k\right]\right) =
  e^{-j\hat{\theta}\left[ k \right]}
  \sum_{i=kN}^{N(k+1)-1}{r\left[i\right] e^{-j\theta\left[i-kN,\hat{\alpha}\left[k\right]\right]}} \notag \\
  &= 
  e^{-j\hat{\theta}\left[ k \right]}
  \sum_{i=kN}^{N(k+1)-1} {\left( s\left[ i \right]e^{j\varepsilon_{\phi}\left[ i \right]} + \eta\left[ i \right] \right) e^{-j\theta\left[i-kN,\hat{\alpha}\left[k\right]\right]}} \notag \\
  &=e^{-j\hat{\theta}\left[ k \right]}
  \sum_{i=kN}^{N(k+1)-1} 
  \left( e^{j\left( \theta\left[ k \right] 
  + \theta\left[ i-kN, \alpha\left[ k \right] \right] 
  + \varepsilon_{\phi}\left[ i \right]\right)} 
  + \eta\left[ i \right] \right) e^{-j\theta\left[i-kN,\hat{\alpha}\left[k\right]\right]} \notag \\
  &=\sum_{i=kN}^{N(k+1)-1}{
  \left( e^{j \varepsilon_{\phi}\left[ i \right]} 
  + \eta\left[ i \right]e^{-j\left(\hat{\theta}\left[ k \right] + \theta\left[i-kN,\hat{\alpha}\left[k\right]\right]\right)} \right)} \notag \\
  &=\sum_{i=kN}^{N(k+1)-1}{e^{j \varepsilon_{\phi}\left[ i \right]}} + \check{\eta}\left[ k \right] \text{,}
\end{align}
where
\begin{align}
  \check{\eta}\left[ k \right] = \sum^{N\left( k+1 \right)-1}_{i=kN}{\eta\left[ i \right]}e^{-j\left(\hat{\theta}\left[ k \right] + \theta\left[i-kN,\hat{\alpha}\left[k\right]\right]\right)}
  \sim
  \mathcal{CN}\left( 0,N \frac{2N_0 B}{T_s} \right)
\end{align}
is a complex Gaussian random variable. Notably, the phase shifts introduced by the matched filter to the original random variable \(\eta\left[ i \right]\) do not alter its statistics \cite{papoulisProbabilityRandomVariables2001}.

Taking the argument of \eqref{eq:g_open}, using a four-quadrant arctangent discriminator (\(\text{atan2}\left\{ \cdot \right\}\)) \cite[Appendix 1]{yoshikawa1990foundations}, we get
\begin{align}
  \label{eq:innovation_full}
  \iota\left[ k \right] &= \text{atan2}\left\{ g \left(\hat{\alpha}\left[k\right],\hat{\theta}\left[k\right]\right) \right\}.
\end{align}

Here, \(\iota\left[ k \right]\), known as innovation, is an error signal that will be used subsequently by the KF algorithm to provide estimates of the carrier phase and Doppler shift and drift. In steady-state conditions, assuming that the power spectral density of the additive thermal noise is sufficiently low, the demodulator ensures that the residual phases \(\varepsilon_{\phi}[i]\) present small values. Hence, \(\text{sin}\left( \varepsilon_{\phi}[i] \right) \approx \varepsilon_{\phi}[i]\) and \(\text{cos}\left( \varepsilon_{\phi}[i] \right) \approx 1\). By substituting these approximations into \eqref{eq:innovation_full}, one can linearize the innovation as
\begin{align}
  \label{eq:innovation_w_assumptions_linearized}
  \iota\left[ k \right] \approx \frac{\text{Im}\left\{ \check{\eta}\left[ k \right] \right\} + \sum_{i=kN}^{N\left( k+1 \right)-1}{\varepsilon_{\phi}[i]}}{\text{Re}\left\{ \check{\eta}\left[ k \right] \right\} + N} 
  \sim \mathcal{N}\left( \frac{1}{N}\sum_{i=kN}^{N\left( k+1 \right)-1}{\varepsilon_{\phi}[i]},\frac{N_0B}{NT_s} \right) \text{.}
\end{align}

With that, it is possible to observe that the innovation statistics carry relevant information regarding the amount of error present on the estimates. Therefore, for small values of \(\varepsilon_{\phi}\left[ m \right]\) and sufficiently high values of \(N\), the usage of the innovations by a linear estimator is justified.

\subsubsection{Bayesian carrier phase tracking system}
\label{subsec:Baysian_carrier_phase_tracking_system}
Now, let us summarize the state-space and measurement models developed in Sections \ref{sec:LOS_model} and \ref{sec:measurement_model} as
\begin{align}
  \label{eq:full_state_meas_models}
  \mathbf{x}\left[ k \right] &= \mathbf{F}\mathbf{x}\left[ k-1 \right] + \mathbf{w}\left[ k-1 \right] \\
  \phi_{LOS}\left[ k \right] &= \mathbf{H}\mathbf{x}\left[ k \right] + v\left[ k \right] \text{,}
\end{align}
where \(\mathbf{x}\left[ k \right] = \big[ \phi_{LOS}\left[ k \right], f_D\left[ k \right], \dot{f}_D\left[ k \right]\big]^\top\) denotes the state space vector, 
\begin{align}
  \mathbf{F} = 
  \begin{bmatrix}
    1 & 2\pi T & \pi T^2 \\
    0 & 1 & T \\
    0 & 0 & 1
  \end{bmatrix}
\end{align}
represents the state transition matrix and \(\mathbf{w}\left[ k-1 \right] \sim \mathcal{N}\left( 0, \mathbf{Q} \right)\) indicates the state noises vector, whose covariance matrix is given by \eqref{eq:state_covariance}. In addition, \(\mathbf{H} = \left[ 1,0,0 \right]\) characterizes the measurement transition matrix and \(v\left[ k \right] \sim \mathcal{N}\left( 0,\sigma^2_v\right)\) is the measurement noise.

Thus, Algorithm \ref{alg:kf} shows the recursive KF equations that provide the optimal linear state and error covariance matrix estimates, given by \cite{brownIntroductionRandomSignals1997}
\begin{align}
  \hat{\mathbf{x}}\left[ k \mid k-1 \right] &= \mathbb{E}\big\{ \mathbf{x}\left[ k \right] \mid \iota\left[ k-1 \right], \iota\left[ k-2 \right], \dots \! \big\}, \\
  \hat{\mathbf{P}}\left[ k \mid k-1 \right] & = \mathbb{E}\big\{ \lVert \mathbf{x}\left[ k \right] - \hat{\mathbf{x}}\left[ k \mid k-1 \right] \rVert^2 \big\} \text{.}
\end{align}
The notation \(\left[ k \mid k-1 \right]\) denotes a sample at an instant \(k\) conditioned to all innovation samples from the initialization up to an instant \(k-1\) and \(\lVert \cdot \rVert\) represents the $L_2$-norm operator.

\begin{algorithm}[H]
    \DontPrintSemicolon
      
      \KwInput{$\hat{\mathbf{x}}\left[ 1 \mid 0 \right], \hat{\mathbf{P}}\left[ 1 \mid 0 \right], \iota\left[ k \right], \mathbf{F}, \mathbf{Q}, \mathbf{H},\sigma^2_v$}
      \KwOutput{$\hat{\mathbf{x}}[k \mid k-1] \ \forall \ k \in \mathbb{Z}$}
    
    \tcp{Update step}
    \If{$k > K$}
    {
        $\mathbf{K}\left[ k \right] \leftarrow \hat{\mathbf{P}}\left[ k \mid k-1 \right] \mathbf{H}\textcolor{black}{^\top} (\mathbf{H} \hat{\mathbf{P}}\left[ k \mid k-1 \right] \mathbf{H}\textcolor{black}{^\top} + \sigma^2_v)^{-1}$
        
        $\hat{\mathbf{x}}\left[ k \mid k \right] \leftarrow \hat{\mathbf{x}}\left[ k \mid k-1 \right] + \mathbf{K}[k]\iota\left[ k \right]$ \label{line:update_x}

        $\hat{\mathbf{P}}\left[ k \mid k \right] \leftarrow \hat{\mathbf{P}}\left[ k \mid k-1 \right] - \mathbf{K}\left[ k \right] \mathbf{H} \hat{\mathbf{P}}\left[ k \mid k-1 \right]$
    }
    \Else{
        $\hat{\mathbf{x}}\left[ k \mid k \right] \leftarrow \hat{\mathbf{x}}\left[ k \mid k-1 \right]$

        $\hat{\mathbf{P}}\left[ k \mid k \right] \leftarrow \hat{\mathbf{P}}\left[ k \mid k-1 \right]$
    }
    \tcp{Project ahead}
    $\hat{\mathbf{x}} \left[ k+1 \mid k \right] \leftarrow \mathbf{F}\hat{\mathbf{x}}\left[ k \mid k \right]$ \label{line:project_x_ahead}

    $\hat{\mathbf{P}} \left[ k+1 \mid k \right] \leftarrow \mathbf{F}\hat{\mathbf{P}}\left[ k \mid k \right]\mathbf{F}\textcolor{black}{^\top} + \mathbf{Q}$

    \(k \leftarrow k+1\)
    \caption{Kalman Filter Algorithm}
    \label{alg:kf}
\end{algorithm}

Furthermore, some caveats need to be discussed. It is well known that the Veterbi algorithm inserts a delay \(K\) to compute the survivor path \cite[Section 8.2]{proakisDigitalCommunications2007}. Therefore, while \(k \leq K\), the KF just projects the state variable ahead using the line \ref{line:project_x_ahead} since the values of the innovation are not available yet. This approximation assumes that, during this initial stage, it is possible to describe the evolution of the states in a deterministic way with \(\mathbf{F}\). This is an imperfect approximation, but it is still plausible provided that the initial estimates are reliable enough and that \(K\) is a few samples. Once the first survivor is computed, the \textcolor{black}{VA} provides \(g \left(\hat{\alpha}\left[k\right],\hat{\theta}\left[k\right]\right)\), from which the innovation \(\iota\left[ k \right]\) is calculated. Note that, for the KF to estimate the LOS carrier phase appropriately, the \textcolor{black}{VA} must provide the correct statistics \(g \left(\hat{\alpha}\left[k\right],\hat{\theta}\left[k\right]\right)\). Since the \textcolor{black}{VA}-KF system forms a DD closed loop, a decreasing signal-to-noise ratio (SNR) is expected to degrade the bit detection performance, which in turn degrades the phase estimation performance as the \textcolor{black}{VA} provides wrong statistics that compromises the Gaussianity assumption of the noise and the linear approximation of the phase discriminator. In this case, a KF is not a feasible estimator anymore, and other nonlinear estimators, such as the particle filter \cite{brownIntroductionRandomSignals1997}, should be considered instead. Additionally, during the acquisition time\footnote{The acquisition time occurs  when the system starts or when it tries to recover the lock again.}, the assumption of a Gaussian distribution for \(\varepsilon_{\phi}[i]\) may not hold. Consequently, the optimality of the KF estimates is not guaranteed during the acquisition. If the KF is initialized with highly imprecise orbital estimates, it may fail to converge, resulting in significant degradation of the demodulator's performance. \textcolor{black}{To circumvent this issue, one can consider using a pilot bit sequence during the acquisition time. When combined with sufficiently accurate initial estimates, this allows the KF to converge efficiently for SNRs of practical concern.} 

Finally, assuming that, within a symbol period \(T\), the LOS phase samples can be approximated as a Taylor series expansion truncated at the second order, we have
\begin{align}
  \label{eq:hat_phi_LOS_n}
  \hat{\phi}_{LOS}\left[ n \right] = & \hat{\phi}_{LOS}\left[ k \mid k-1 \right] \nonumber \\
  & + 2\pi \hat{f}_D\left[ k \mid k-1 \right](n - kN)T_s \nonumber \\
  & + \pi \hat{\dot{f}}_D \left[ k \mid k-1 \right] \left( (n - kN)T_s \right)^2 \text{,}
\end{align}
where
\begin{align}
    \label{eq:hat_phi_LOS_0}
    \hat{\phi}_{LOS}\left[ 1 \mid 0 \right] \triangleq - \frac{2\pi f_{RF}\rho_{s,g}}{c}
\end{align}
is the initial phase estimate \cite{fohlmeisterGNSSCarrierPhase2021}, which is computed only once during the acquisition time by using the true range between the satellite and the ground station, \(\rho_{s,g}\). Here, the symbol \(c \triangleq 299\; 10^6\) m/s denotes the speed of light.

The phase estimates are used by a carrier generator to produce carrier replica samples which are then multiplied to the received signal \eqref{eq:r_tilde_n}, as described in Section \ref{subsec:LOS_correction}. The output of this system is given by
\begin{align}
  d\left[ n \right] = e^{-j \hat{\phi}_{LOS}\left[ n \right]}.
\end{align}
In addition, the delay estimates, that input the Farrow interpolator to correct the LOS delay, are also obtained from \(\hat{\phi}_{LOS} \left[ n \right]\) as follows
\begin{align}
    \label{eq:hat_tau_LOS_n}
  \hat{\tau}_{LOS}\left[ n \right] = \frac{-\hat{\phi}_{LOS}\left[ n \right]}{2\pi f_{RF}} \text{.}
\end{align}

\textcolor{black}{Given that the timing delay estimates depend on the carrier phase estimates provided by the KF proposed in this work, it is important to highlight that they are susceptible to ambiguities. Thus, if the initial carrier phase considered by the KF differs significantly from the expected LOS phase determined by the receiver-satellite range, a constant bias in the delay estimates is likely to appear.}

\textcolor{black}{An error in the delay synchronization would directly affect the matched filter output, which is given by \eqref{eq:v_kk}. It is worth noting that \eqref{eq:g_k_k} is only an approximation for the case where timing delay synchronization is perfect. In this context, the matched filter outputs would be corrupted by the residual delay present in \eqref{eq:r_corrected}. Therefore, a reduction of BER performance is expected when the delay bias is large.}

\textcolor{black}{Additionally, since the KF relies on the correctness of the estimated bits provided by the VA, an increase in the variance of its estimates is anticipated. A complete evaluation of the behaviour of the proposed demodulator under severe discrepancies between the true and estimated delay is beyond the scope of this work. Nevertheless, in the beginning of Section \ref{sec:results}, we provide preliminary heuristics on how precise the initial values of the phase should be in order to avoid large errors on the delay estimates.}

\begin{figure}[H]
  \centering
  \includegraphics[scale=1]{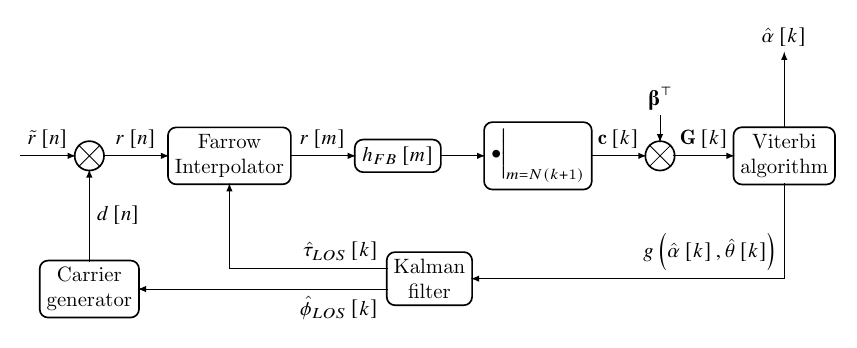}
  \caption{Complete AFSK demodulator architecture, including the Viterbi detector and the KF algorithm.}
  \label{fig:full_demodulator}
\end{figure}

The full demodulator architecture is represented in Figure \ref{fig:full_demodulator}.

%% file: secs/results.tex
\section{Numerical results} \label{sec:results}
In this section, performance results, obtained through realistic computer simulation, of the proposed demodulator are presented and discussed. The system performance is assessed in terms of the BER for different signal-to-noise ratio (SNR) scenarios, with LOS phase and timing impairments computed from real CubeSat orbits. The quality of the KF estimates are analyzed based on a MRMSE statistic, with a window size corresponding to the amount of samples within 3 seconds of simulation time (3600 symbol periods). The reasoning behind choosing a moving window statistic metric instead of a static full window statistic is related to the possibility of assessing the quality of the KF estimates over time of each simulated CubeSat LOS dynamics scenarios. In addition, we also present the initial convergence time of the trace of the KF error covariance matrix estimates, as a metric of reliability of the state estimates. In \cite{tapyumas3:online}, the authors made the MATLAB\textsuperscript{\textcopyright} source code of the model used to generate the results in this section freely available and open-source to the public.

The following parameters were used to simulate an AFSK uplink communication scenario for CubeSats:
\begin{itemize}
    \item \textit{General parameters}
    \begin{itemize}
        \item \textcolor{black}{Uplink} carrier frequency, \(f_{RF} = 149\) MHz;
        \item Bit rate, \(R = 1.2\)kbps;
        \item Sampling frequency, \(f_s = 480\) kHz;
        \item Difference between the frequencies used to modulate the data, \(\Delta f = 1000\) Hz;
        \item Modulation index, \(h = 2 \Delta f / R = 5/6\);
    \end{itemize}
    \item \textit{\textcolor{black}{VA} and KF parameters}
    \begin{itemize}
        \item \textcolor{black}{VA} delay, \(K=25\)\textcolor{black}{. In order to compute the MLSD metric, the VA needs to observe \(K\) received symbols before confidently deciding on the first symbol of the detected path through the trellis. After this initial stage, each subsequent step of the VA is performed recursively by updating each node's memory with the transition metrics, and then selecting the path corresponding to the metric with the highest value (i.e., the survivor path). Although the VA computes the maximum likelihood recursively, which guarantees convergence in the long run, a high value of \(K\) ensures that the first symbols are correctly decided---this is a critical step as these symbols are used during the acquisition time. If they are incorrect, the acquisition step may fail, preventing the tracking module from locking onto the received phase---however, a high value of \(K\) delays the KF's ability to adapt to changes in the environment as it processes delayed observations of the incoming signal. This trade-off is addressed by selecting the minimum value of \(K\) that ensures reliable initial detected symbols. The approach adopted in this article is the path memory truncation, which is based on the signal memory, defined\footnote{This definition is equivalent to determining the minimum number of transitions required for the finite state machine to completely forget the memory of its current state.} as the number of transitions needed for the trellis to reach the steady state starting from a given node \cite[p. 244]{proakisDigitalCommunications2007}. In our case, a binary and full-response signaling with 12 states, the AFSK trellis reaches its steady stage after the fifth transition \cite{pacelliAlldigitalAFSKModem2020}. Empirical analysis indicates that \(K\) should be, at least \(5\) times the signal memory in order to guarantee a high probability of convergence of all survivor into a unique path through the trellis \cite[p. 246]{proakisDigitalCommunications2007}. Therefore, the value of \(K\) is set to \(25\). Additionally, it is important to mention that the present article makes use of 1200 pilot bits to initialize the system. During this phase, the \textcolor{black}{VA} does not work in the sense that no metrics are computed (only the phase memory \(\hat{\theta} \left[ k \right]\) is updated). Instead, the metrics \(g \left(\hat{\alpha}\left[k\right],\hat{\theta}\left[k\right]\right)\) of the known bits are delivered to the KF, which uses them to start the acquisition process. After this short period, the VA starts working, finishing the acquisition process or tracking the signal, depending on the scenario.}
        \item The state-space noise variances modeling the uncertainty of the LOS dynamics and the clock drift were chosen as \(\sigma^2_1 = \sigma^2_2 = 0\) and \(\sigma_3^2 = 0.01\). \textcolor{black}{Hence, we assume that the clock drift is negligible and that only the uncertainty about the LOS dynamics is relevant \cite[Section 2.4.1]{fohlmeisterDualKalmanFiltering2018}. The rationale behind choosing the state-space noise variances is that, for a fixed value of the measurement noise variance, higher state-space noise variances make the filter more responsive to sources of impairments, such as increased satellite-receiver dynamics, as well as ionospheric and tropospheric effects. However, despite the increasement of system responsiveness, the residual noise of the KF estimates also increases. In other words, the configured state and measurement noise variances of a KF-PLL behave anagously to the noise bandwidth of the loop filter in a conventional PLL or FLL \cite{OnTheIdentifiabilityOfCovariances}. Since the performance of the VA depends on the variance of the the phase error \eqref{eq:phase_error}, we can expect its performance to degrade for very high state-space noise variances. Therefore, we emphasize that the state-space noise variances should be selected according to the expected sources of impairments in each application. In the case of CubeSat applications, simulations have shown that the chosen values in this work are suitable for tracking its high LOS dynamics profile, while providing a sufficiently low residual noise so that the BER is minimally affected;}
        \item Following the idea of using common SNR estimators \cite{MeasurementNoiseEstimators} provided by standard receivers, as discussed in \cite{OnTheIdentifiabilityOfCovariances}, and assuming that the power of the received signal is normalized to one, one can use the estimate of the measurement noise variance as \(\hat{\sigma}^2_v = 10^{\frac{-SNR_{dB}}{10}}\). In the following simulations it is assumed that the ground base station is automatically adapting the power level of the transmitted signal, in order to fix the SNR level throughout the simulation time;
        \item The initial state \(\mathbf{\hat{x}}\left[ 1 \mid 0 \right] = \big[ \hat{\phi}_{LOS}\left[ 1 \mid 0 \right], \hat{f}_D\left[ 1 \mid 0 \right], \hat{\dot{f}}_D\left[ 1 \mid 0 \right] \big]^\top \) can be derived from the satellite-station true range provided by the model. On the one hand, the LOS phase is computed as \eqref{eq:hat_phi_LOS_0}. On the other hand, the initial Doppler shift and drift are computed by employing numerical differentiation techniques\footnote{This can be easily accomplished by estimating the satellite trajectory around the TLE (two-line element) epoch.}. The reader should note that, at the KF initialization, \(\hat{\phi}_{LOS}\left[ 1 \mid 0 \right]\) is unwrapped and directly related to relative distance between the satellite and the ground station (\textit{cf.} \eqref{eq:hat_phi_LOS_0}). \textcolor{black}{If the initial range is imperfect, after the phase correction, a residual LOS phase tracking error remains in the system. However, this unwrapped phase error is lost due to the arc tangent discriminator operation, which makes it insensitive to phase offsets greater than \(\left[ -\pi, \pi \right]\). The integer number of phase cycles left in the system biases the estimated delay, as shown in \eqref{eq:hat_tau_LOS_n}. The satellite orbital model used to generate \(\rho_{s,g}\) is the SGP4 algorithm \cite{SGP4OrbitDetermination}---for our system configuration, this algorithm must be accurate enough so that the initial range error is less than\footnote{By applying this range in \eqref{eq:hat_phi_LOS_0} and converting the phase into group delay using \eqref{eq:hat_tau_LOS_n}, we obtain a delay offset which is close to a sample period for the proposed demodulator.} \(608\) meters. Otherwise, the integer factor of phase ambiguity causes \(\hat{\tau}_{LOS}\left[ n \right]\) to be biased by more than one sample, resulting in a subtle mismatch in the filter bank. However, as the initial range error increases, a BER degradation is expected. For instance, in a scenario where the initial range error leads to a timing delay offset of \(50\) samples, at \(8\) dB, the BER increases from \(6.9 \times 10^{-5}\) to \(2.7 \times 10^{-4}\);}
        \item The initial covariance error matrix \(\mathbf{P}\left[ 1 \mid 0 \right]\) is configured as a diagonal matrix with element values \(\left\{ \pi^2/3, \left( 0.1 \right)^2/12, \left( 0.01 \right)^2/12 \right\}\), respectively. These values were chosen as the variances of the uniform distributions \(\hat{\phi}_{LOS}\left[ 1 \mid 0 \right] \sim \mathcal{U}\left( -\pi,\pi \right)\) rad, \(\hat{f}_D\left[ 1 \mid 0 \right] \sim \mathcal{U}\left( -0.05, 0.05 \right)\) Hz, and \(\hat{\dot{f}}_D\left[ 1 \mid 0 \right] \sim \mathcal{U}\left( -0.005, 0.005 \right) \) Hz/s for initial phase, Doppler shift, and Doppler drift, respectively, which corresponds to the assumed maximum errors related to the SGP4 orbit estimation. It is important to comment that the initial covariance matrix values affects the convergence time of the KF. With that, for an initial covariance matrix with larger values, the estimates of the filter are expected to converge rapidly. However, if its initial covariance values are too large, it is possible to induce the filter to divergence. Thus, it is necessary to evaluate which are the best initial covariance matrix values for the desired application;
     \end{itemize}
\end{itemize}

\subsection{Satellite orbits and LOS effects}

The satellite orbits are derived from scintpy, a Python package developed by the authors and made freely available and open source in \cite{tapyusci47:online}. This package searches for currently active CubeSat IDs on \url{https://celestrak.org}. Based on the collected NORAD IDs of the active CubeSats, a set of TLEs (two-line elements \cite{RevisitingSpaceTrackReport3}) are downloaded from \url{https://www.space-track.org}. Each CubeSat contains a TLE (two-line element) whose epoch is the closest to the desired UTC date time (in our case, it is set to 28th October 2024, 07:00:00 AM). Then, it is searched for all satellites that, for this date time, are in line-of-sight with the ground station location, which is set to São José dos Campos, São Paulo, Brazil, at the geographic coordinates \(23.2198^\circ\) S, \(45.8916^\circ\) W. The same process is repeated in intervals of 30 minutes until a total of 10 CubeSats are found for the same fixed location. Table \ref{tab:satellites} shows the NORAD identification, rise and set time, and satellite name for all CubeSats used in the simulations. The skyplot and the LOS effects for each CubeSat are shown in Figure \ref{fig:main-orbits}.

\begin{table}[H]
    \centering
    \begin{tabular}{|l|c|c|c|}
        \hline
        \textbf{Satellite Name} & \textbf{NORAD ID} & \textbf{Rise Time (UTC)} & \textbf{Set time (UTC)} \\ \hline
        FUNCUBE 1 & 39444 & 06:57:12 & 07:00:49 \\ \hline
        TIGRISAT & 40043 & 06:54:21 & 07:04:45 \\ \hline
        SITRO-AIS-52 & 59063 & 06:59:17 & 07:04:00 \\ \hline 
        POPACS 2 & 39269 & 07:29:02 & 07:35:36 \\ \hline
        POLYITAN 1 & 40042 & 07:23:06 & 07:30:25 \\ \hline
        M2 PATHFINDER & 45727 & 07:27:14 & 07:32:21 \\ \hline
        CUTE-1 & 27844 & 07:59:56 & 08:11:29 \\ \hline
        AEROCUBE 8D & 41852 & 07:59:56 & 08:08:07 \\ \hline
        AEROCUBE 8C & 41853 & 08:21:51 & 08:31:17 \\ \hline
        DUCHIFAT-3 & 41853 & 08:21:44 & 08:31:29 \\ \hline 
    \end{tabular}
    \caption{The satellite name, NORAD identification, and its respective rise and set time.}
    \label{tab:satellites}
\end{table}

\begin{figure}[H]
    \centering
    \begin{subfigure}{0.6\textwidth}
        \includegraphics[width=\linewidth]{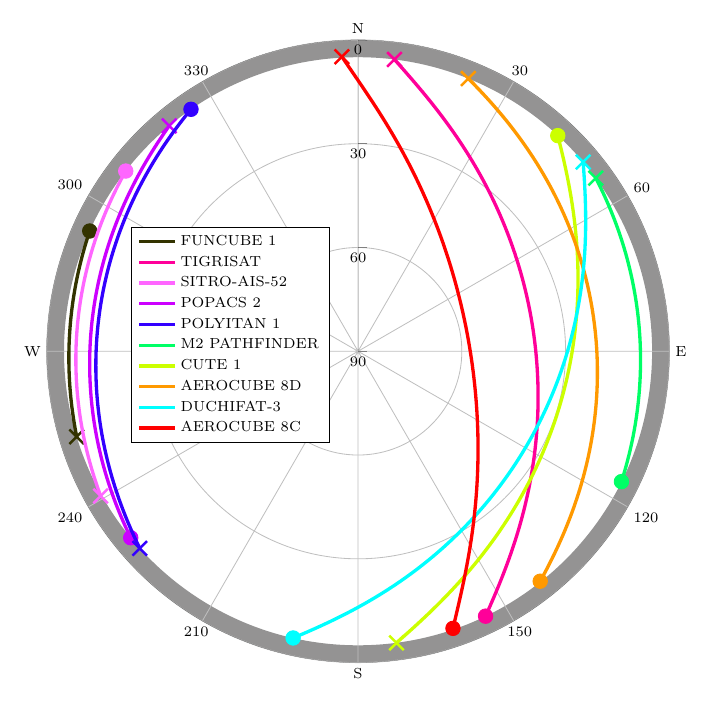}
        \subcaption{The skyplot of the CubeSats used in the simulations. In this figure, the radius indicates the elevation angle, where \(90^\circ\) is the zenith. The gray area is the elevation angle below \(5^\circ\), which is considered the out of the line of sight. The circles and crosses indicate the starting and ending points, respectively.}
        \label{subfig:skyplot}
    \end{subfigure}
\end{figure}
\begin{figure}[H]\ContinuedFloat
    \centering
    \begin{subfigure}{0.6\textwidth}
        \includegraphics[width=\linewidth]{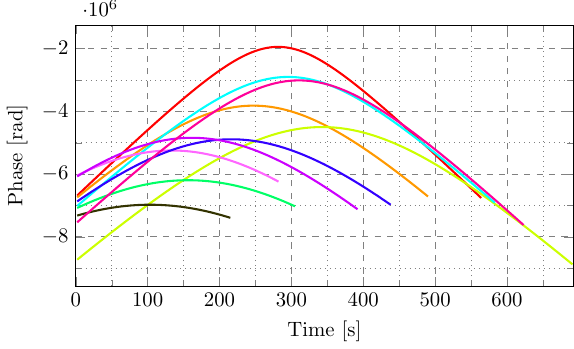}
        \subcaption{The LOS phase for each CubeSat.}
        \label{subfig:phase}
    \end{subfigure}
\end{figure}%
\begin{figure}[H]\ContinuedFloat
    \centering
    \begin{subfigure}{0.6\textwidth}
        \includegraphics[width=\linewidth]{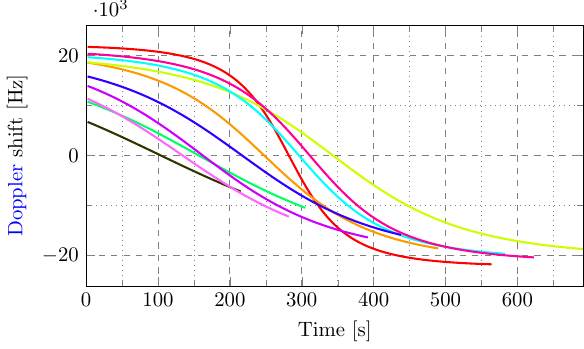}
        \subcaption{The Doppler shift for each CubeSat.}
        \label{subfig:fD}
    \end{subfigure}
\end{figure}%
\begin{figure}[H]\ContinuedFloat
    \centering
    \begin{subfigure}{0.6\textwidth}
        \includegraphics[width=\linewidth]{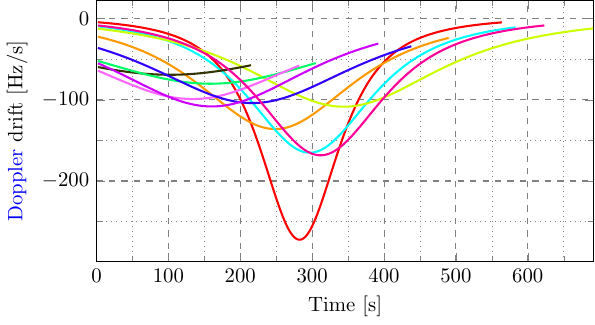}
        \subcaption{The Doppler drift for each CubeSat.}
        \label{subfig:fD_dot}
    \end{subfigure}
    \caption[]{The skyplot and the LOS phase, Doppler shift, and Doppler drift for each CubeSat. The starting instant of all LOS effects is set to 0 for the sake of readability.}
    \label{fig:main-orbits}
\end{figure}

\subsection{Bit error rate (BER)}
\begin{figure}[H]
    \centering
    \includegraphics[scale=0.9]{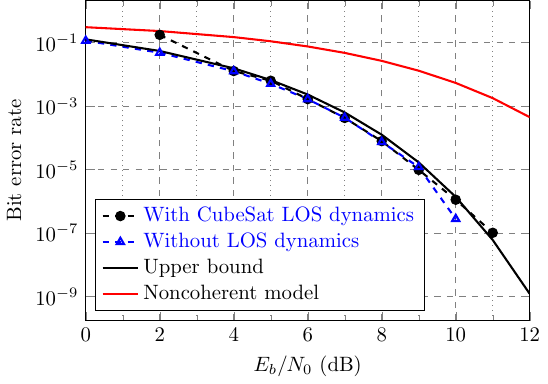}
    \caption{\textcolor{black}{Comparison of the BER for the proposed model considering CubeSat LOS dynamics and with no LOS dynamics, the} theoretical probability error upper bound for coherent AFSK, and the theoretical probability error for the noncoherent AFSK demodulator.}
    \label{fig:BER}
\end{figure}

Figure \ref{fig:BER} shows the obtained system BER performance, \textcolor{black}{for the cases with the CubeSat LOS dynamics and without any LOS dynamics (i.e. a hypothetic static receiver-transmitter relative range scenario),} where the value for each \(E_b/N_0\) is an average of 100 Monte Carlo simulations in total\footnote{This is the result of 10 independent Monte Carlo runs per CubeSat. By considering the observation window time (set minus rise time) of each satellite, a total of 12 hours, 46 minutes, and 30 seconds was simulated.}. Each Monte Carlo run is simulated with randomized seed for channel noise and bit stream. For the sake of comparison, the plot also shows the theoretical probability error for the noncoherent AFSK demodulator, which can be found in \cite{mengaliSynchronizationTechniquesDigital1997}, and \textcolor{black}{the} theoretical probability error upper bound \textcolor{black}{for a coherent AFSK demodulator}, as derived in \cite{proakisDigitalCommunications2007}. 

\textcolor{black}{By comparing the coherent and the noncoherent theoretical BER behavior, we may expect that the proposed model significantly outperform the standard noncoherent AFSK demodulator for at least \(5\) dB. This supposition is confirmed by comparing the BER performance for the static case (i.e., no LOS dynamics) with the theoretical coherent upper bound. Furthermore, we argue here that the proposed system is well-suited for CubeSat applications by observing that there is only a slight performance degradation between the dynamic and the static scenarios for $E_b/N_0 \leq 9$ dB. It is worthwhile to note that the increased difference between the BER for the dynamic and static scenarios in the cases when $E_b/N_0 > 9$ is mostly dominated by LOS-induced errors, instead of noise. The performance of the proposed KF for higher Wiener model orders should be assessed in future works. For dynamic and static scenarios, we obtained no bit errors for $E_b/N_0 > 11$ dB and $E_b/N_0 > 10$ dB, respectively. In contrast, for \(E_b/N_0 < 2\) the system cannot lock onto the received signal, and therefore no BER is obtained.}

\textcolor{black}{It is important to mention here that, although the literature reports performance of coherent and noncoherent FSK demodulators, they are not directly comparable to the proposed model due to two main issues: 1) due to its application, Doppler frequency profile experienced by the demodulator in this work is very severe, reaching at \(20\) kHz in the most critical cases. In contrast, for other applications, such a IoT, the Doppler frequency profile is usually much lower, reaching at most \(2\) kHz \cite{Lim2024SDRIoT}; 2) the demodulator performance is heavily dependent on the modulation index, which is \(5/6\) for the AFSK modem. Therefore, a proper comparison should only be made with FSK-based modems with same modulation index. At the best of our knowledge, there is no work that reports the performance of a coherent FSK demodulator with the same modulation index as the proposed model.}

\textcolor{black}{
    \subsection{Loss-of-lock occurrence}
    Due to the high Doppler frequency profile experienced by CubeSats, signal tracking is a major challenge of the proposed demodulator, which may lose lock on the received signal for low SNR. Figure \ref{fig:heatmap_loss_of_lock} presents the heatmap of the loss-of-lock occurrence for each \(E_b/N_0\) and satellite combination. For \(E_b/N_0 < 2\) dB, the proposed demodulator is not operational, as the tracking module cannot lock onto the received signal. However, for \(E_b/N_0 \geq 4\) dB, the tracking module successfully locks onto the received signal for most Monte Carlo runs, allowing the system to operate properly. For CubeSat communication systems, this minimum \(E_b/N_0\) value for proper operation is considered acceptable, as for some modulation schemes, the available \(E_b/N_0\) before applying any error correction codes is predicted to be around 9 dB for BER values of practical interest \cite{aragonResearchingRobustCommunication2011,pwsat2teamPhaseDocumentationCritical2016} \cite[Section 5.3]{sebestyenLowEarthOrbit2018}.
    \begin{figure}[H]
        \centering
        \includegraphics[scale=0.9]{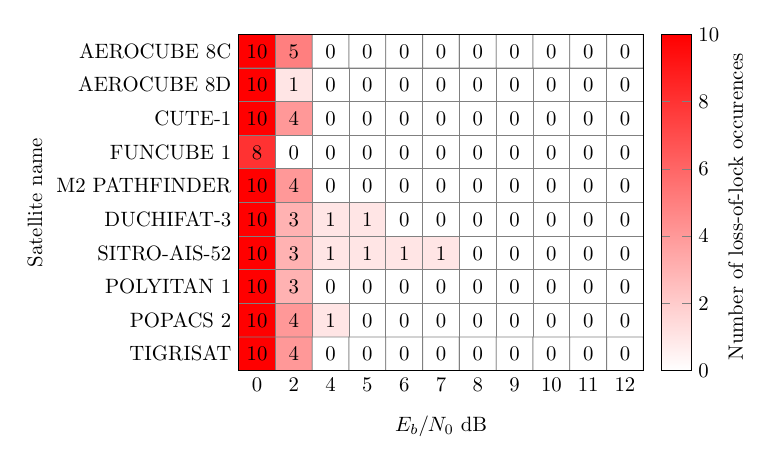}
        \caption{\textcolor{black}{Heatmap indicating the number of Monte Carlo runs in which a loss of lock occurs. A total of 10 Monte Carlo runs were performed for each CubeSat and \(E_b/N_0\) combination.}}
        \label{fig:heatmap_loss_of_lock}
    \end{figure}
}

\subsection{Kalman filter performance}

Figure \ref{fig:main-mrmse} shows the obtained MRMSE statistic for the Doppler shift \(\hat{f}_D\left[ k \mid k-1 \right]\) and drift \(\hat{\dot{f}}_D \left[ k \mid k-1 \right]\) estimates provided by the KF algorithm, for one Monte Carlo simulation, for each CubeSat, and for \(E_b/N_0 = 8\) dB. Comparing this plot with Figure \ref{subfig:fD} and \ref{sub@subfig:fD_dot}, it is possible to observe that the MRMSE of \(\hat{f}_D\left[ k \mid k-1 \right]\) and \(\hat{\dot{f}}_D \left[ k \mid k-1 \right]\) increases according to the rate change of the Doppler frequency shift and drift, respectively. Although the estimates are slightly more imprecise for the intervals of high rate of change, it does not significantly impact the BER, as can be inferred from Figure \ref{fig:BER}.
\begin{figure}[H]
    \centering
    \begin{subfigure}{0.7\textwidth}
        \includegraphics[width=\linewidth]{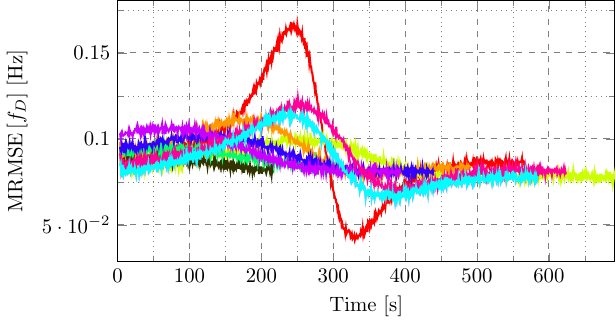}
        \subcaption{MRMSE of the Doppler shift estimate.}
        \label{subfig:mrmse-fD}
    \end{subfigure}
\end{figure}
\begin{figure}[H]\ContinuedFloat
    \centering
    \begin{subfigure}{0.7\textwidth}
        \includegraphics[width=\linewidth]{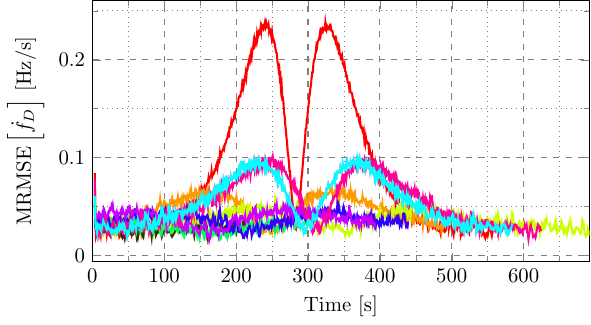}
        \subcaption{MRMSE of the Doppler drift estimate.}
        \label{subfig:rmse-fD_dot}
    \end{subfigure}
    \caption[]{MRMSE for all CubeSats. The notation \(\textnormal{MRMSE}\left[ \cdot \right]\) denotes the moving window root mean squared error operator. Once again, the starting time of all MRMSE curves is set to 0 for the sake of readability.}
    \label{fig:main-mrmse}
\end{figure}%

\begin{figure}[H]
    \centering
    \includegraphics[scale=0.9]{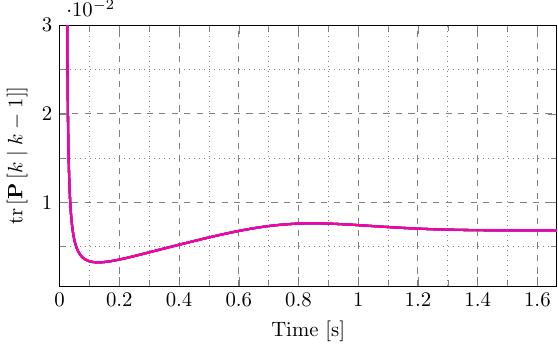}
    \caption{Error covariance matrix trace, for 1 Monte Carlo run and each studied LOS dynamics scenario. In this figure, \(\textnormal{tr} \left[ \cdot \right]\) denotes the trace operator. The starting time of all trace curves is set to 0 for the sake of readability.}
    \label{fig:error_cov_trace}
\end{figure}

In sequence, Figure \ref{fig:error_cov_trace} shows the convergence of the trace of the KF error covariance matrix estimates \(\mathbf{P}\left[ k \mid k-1 \right]\) in the initialization of the algorithm, for one Monte Carlo simulation, for each CubeSat, and for \(E_b/N_0 = 8\) dB. This plot shows that the trace of the estimated error covariance matrix successfully have the same convergence behavior for all CubeSats, showing that the state estimates have converged and are reliable.

%% file: secs/conclusion.tex
\section{Conclusion} \label{sec:conclusions}
In this work, the design of a new all-digital and coherent AFSK demodulator suitable for TT\&C modules for CubeSats was presented. The proposed DD AFSK demodulator architecture, which is based on \textcolor{black}{VA} for bit detection and a KF for synchronization was derived and detailed. The \textcolor{black}{VA-KF} form a closed-loop and DD system which provides coherent bit detection and joint phase and timing synchronization for CubeSat LOS dynamics.

The proposed demodulator is subject to additive white Gaussian noise channel with real CubeSat orbits in LEO, where the ground station is set to São José dos Campos, Brazil. The performance results obtained through computational simulations demonstrate that the proposed model can withstand such scenarios with a gain of 5 dB in terms of BER compared to the conventional noncoherent AFSK demodulator. Moreover, when compared to \textcolor{black}{a hypothetical} static scenario (no LOS dynamics), the proposed demodulator has no BER degradation, except for very high values of \(E_b/N_0\), where the LOS-induced bit errors dominate. Finally, the quality of the KF estimates was evaluated through a MRMSE statistic and the trace of the estimated error covariance matrix. \textcolor{black}{The BER advantage of the proposed joint synchronization and demodulation scheme, however, comes at the cost of higher algorithmic complexity and the requirement of 1200 pilot bits during acquisition.}

The proposed joint synchronization and bit detection algorithm provides excellent performance under realistic CubeSat scenarios, significantly outperforming the standard noncoherent AFSK demodulator, while introducing only a moderate complexity increase in the satellite receiver. \textcolor{black}{Future efforts might consider embedding the proposed demodulator into an FPGA circuit board or using a software-defined radio for real validation. All components of the AFSK demodulator might initially be implemented through a design methodology based on circuit synthesis, which allows fast prototyping of the logical modules.}